\documentstyle[sprocl,epsf]{article}

\def\Journal#1#2#3#4{{#1} {\bf #2}, #3 (#4)}

\def\NPB{{\em Nucl. Phys.} B}
\def\PLB{{\em Phys. Lett.}  B}
\def\PRL{\em Phys. Rev. Lett.}
\def\PRD{{\em Phys. Rev.} D}
\def\ZPC{{\em Z. Phys.} C}
\def\JETP{{\em Sov. Phys. JETP}}

\def\Pom{{\bf I\!P}}

\def\be{\begin{equation}}
\def\ee{\end{equation}}
\def\bea{\begin{eqnarray}}
\def\eea{\end{eqnarray}}

\begin{document}

\title{THE TENSOR STRUCTURE FUNCTION $b_2(x,Q^2)$ OF THE DEUTERON AT
SMALL $x$}

\author{W. SCH\"AFER}

\address{Institut f\"ur Kernphysik, 
Forschungszentrum J\"ulich, D-52425 J\"ulich} 


\maketitle\abstracts{ The dependence of the diffractive nuclear shadowing and the 
nuclear pion excess effect on the deuteron spin alignment leads to a substantial 
 tensor polarisation of sea partons in the deuteron. The corresponding tensor structure function 
$b_2(x,Q^2)$ rises towards small $x$, and we predict the about one percent tensor asymmetry
$A_2(x,Q^2) =b_2(x,Q^2)/F_{2D}(x,Q^2)$. We calculate the shadowing of the gluon distribution
in the deuteron for two different pomeron gluon-distributions. 
For a hard gluon dominated $\Pom$-structure function, as advocated by H1, we find
large gluon shadowing, which appears to be in conflict with 
the DGLAP analysis of nuclear SF's.}

\section{The shadowing driven tensor structure function}
It is a common theoretical prejudice, that in the high energy limit ($x \to 0$ in
DIS), total cross sections cease to depend upon the beam and target polarizations.
However we shall see in the following, that the tensor polarization of sea quarks in
a deuteron is substantial and even rises for $x \to 0$ \cite{tensor}.  Let $\vec{S}$
be a spin-1 operator, then the {\it{tensor polarization}} $T_{ik} = \frac{1}{2}( S_i
S_k + S_k S_i - \frac{2}{3} \vec{S}^2 \delta_{ik} )$ is a T-even and P-even
observable. As such it is accessable in DIS of {\it{unpolarized}} leptons on the
{\it{(tensor-)polarized}} deuteron.  Let $\vec{n}$ be the deuteron-spin quantization
axis, and $\lambda = \vec{S}\cdot \vec{n}$ the spin projection upon the quantization
axis. Then the total photoabsorption cross sections $\sigma^{(\lambda
  )}(\gamma^*_{T,L} D)$ bear a dependence upon the deuteron polarization state, and
thus on the tensor polarization \cite{Hood} (for the case of real photons see
also\cite{Pais}): $\sigma^{(\lambda )}(\gamma^*_{T,L} D) = \sigma_0 + \sigma_2 T_{ik}
n_i n_k + ...$ The corresponding structure functions are introduced through \be
F_2^{(\lambda )} = \frac{Q^2}{4 \pi^2 \alpha_{em}}(\sigma^{(\lambda )}(\gamma^*_{T}
D) + \sigma^{(\lambda )}(\gamma^*_{L} D)).  \ee If one choses the $\gamma^* - D
$-collision axis for the quantization axis ($z-$ axis), the structure function $b_2$
can be defined as: \bea b_2 (x,Q^2) &=& \frac{1}{2} [ F_2^{(+)}(x,Q^2)+
F_2^{(-)}(x,Q^2)- 2 F_2^{(0)}(x,Q^2) ] \nonumber \\ &=& \frac{1}{2} \sum_{f} e_f^2 x[
q_f^{(+)}(x,Q^2)+ q_f^{(-)}(x,Q^2)- 2 q_f^{(0)}(x,Q^2) ].  \eea Here the last line is
the expression for $b_2$ in the parton model, and $q_f^{(\lambda )}$ stands for the
parton density in the hadron with spin projection $\lambda$ on the $z-$axis.  In the
same manner one defines the transverse $b_T = 2xb_1$ and longitudinal $b_L = b_2 -
b_T$ structure functions.
\begin{figure}[h]
\epsfxsize=1.0\hsize
\epsfbox{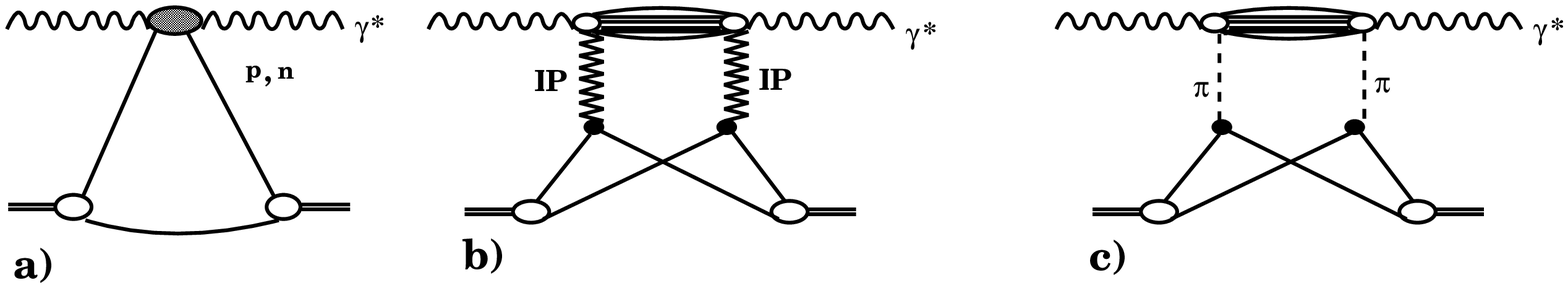}
\caption{Impulse approximation (a) and Gribov's inelastic shadowing (b),(c) diagrams for DIS
on the deuteron.}
 \label{fig3}
\end{figure}
It was only the impulse approximation (IA) that had been employed in the discussion
of the tensor structure functions before \cite{Hood,Umn}. What has been found is a
contribution to $b_2$ that vanishes in the limit $x \to 0$ and is only the per-mille
effect even at moderate $x$.  Furthermore, in the IA one can derive the sum rule $
\int_0^1 dx~b_1(x,Q^2) = 0. $ A conjecture had been made by Close and Kumano \cite{Close}, 
for it to hold also beyond the impulse approximation, under the assumption of an
unpolarised sea. We shall see that the results of the beyond-the-IA evaluation of the
tensor structure function imply that such a sum rule for $b_1$ cannot exist.  The
appropriate framework for DIS on nuclear targets is the Glauber-Gribov multiple
scattering theory\cite{Gribov,NZ1}. The inelastic shadowing diagrams lead to the
nuclear eclipse effect: $\sigma_{tot}(\gamma^* D) = \sigma_{tot}(\gamma^* p) +
\sigma_{tot}(\gamma^* n) - \Delta \sigma_{sh}(\gamma^* D)$ and two production mechanisms
of the inelastically exited intermediate state $X$ have to be distinguished: in
diffractive nuclear shadowing (NSH) the state $X$ is produced by pomeron exchange,
the second mechanism is the exitation of $X$ by pion exchange.  The diffractive NSH
contribution can readily be calculated using Gribov's theory extended to DIS
in\cite{NZ1}: \bea \Delta F^{(\lambda )}_{sh}(x,Q^2) = \left.
\frac{2}{\pi}\frac{Q^2}{4 \pi^2 \alpha_em} \int d^2\vec{k}_\perp \int dM^2
S^{(\lambda)}_D(4 \vec{k}^2) {d\sigma^D(\gamma^* \to X) \over dt
  dM^2}\right|_{t=-\vec{k}^2}.  \eea Here $M$ is the mass of the state $X$, $\vec{k}
= (\vec{k}_\perp , k_z)$ is the momentum of the pomeron, and $S^{(\lambda)}_D(4
\vec{k}^2) = \int d^3 \vec{r} |\Psi^{(\lambda)}_D(\vec{r})|^2
exp(-i\vec{r}\cdot\vec{k})$ denotes the deuteron formfactor. It is of crucial
importance to include in the calculation the $D-$wave admixture in the deuteron
wavefunction $\Psi^{(\lambda)}_D$, as it turns out that the diffractive NSH
contribution to the tensor structure function is mainly due to the $S-D-$
interference. A further ingredient of the calculation is the differential cross
section $ d\sigma^D(\gamma^* \to X) / dt dM^2$ for the diffraction dissociation
of the virtual photon into the state $X$ on the proton, which can be written in terms
of the diffractive structure function.  We make use of a theoretical calculation of
the diffractive cross section \cite{GNZ} that reproduces the HERA data on diffraction
-- this should be demanded of any reliable calculation of the nuclear shadowing.
Schematically, the diffractive contribution can be cast into the convolution form:
\be \Delta F^{(\lambda )}_{sh}(x,Q^2) = \int_x^1 {dx_\Pom \over x_\Pom} \Delta
n_{\Pom/D}^{(\lambda)}(x_\Pom) F_{2\Pom}({x \over x_\Pom },Q^2), \ee
where $ \Delta n_{\Pom/D}^{(\lambda)}(x_\Pom)$ is the nuclear modification of the
pomeron flux. 
The second mechanism is pion exchange:
it dominates $\gamma^* p \to nX$, and contributes substantially to $\gamma^* p \to
pX$ at moderately small $x_\Pom$. One can derive an analogous convolution 
formula as for the pomeron contribution, only the pomeron flux and structure function
get replaced by the corresponding pionic quantities.
From these representations one can read off the evolution properties of
the two contributions: diffractive NSH is known to be the leading twist, DGLAP-evolving
 effect \cite{NZ1}. That the pionic contribution is DGLAP evolving is
self-evident, as the $Q^2$-dependence enters only through the pion structure function. 
Our results for $b_2$ and the related asymmetry $A_2=b_2/F_{2D}$ can be seen in fig.2a,b) .
\begin{figure}[h]
\epsfxsize=1.0\hsize
\epsfbox{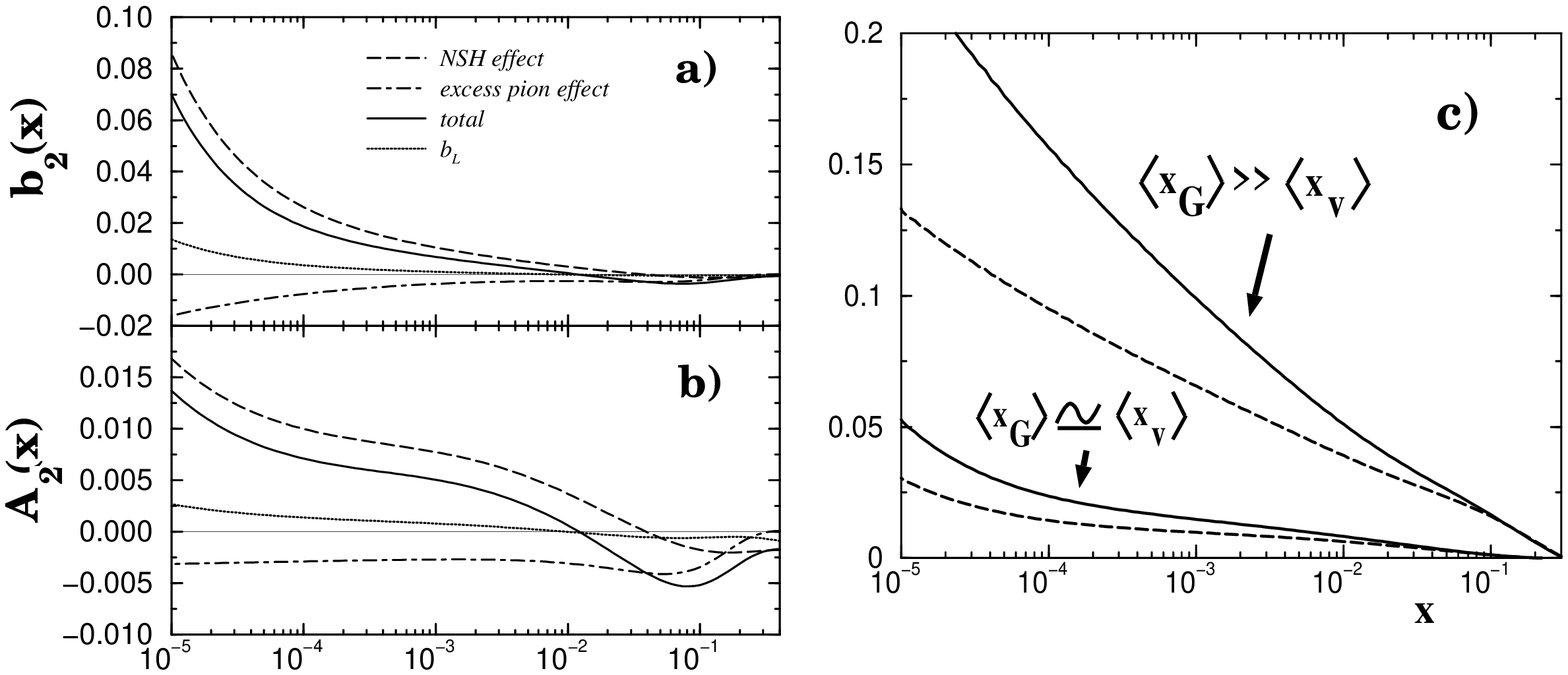}
\caption{(a) The tensor structure function of the deuteron $b_2(x,Q^2)$ at
$Q^2=10 GeV^2$ and its diffractive NSH and pion excess components.
(b) the tensor polarization $A_2(x,Q^2)$ of quarks and antiquarks
in the deuteron at $Q^2=10 GeV^2$. Also shown (by the dotted line) is the 
longitudinal tensor structure function $b_L(x,Q^2)$ (in (a)) and the corresponding
asymmetry (in(b)).(c) Diffractive shadowing correction to the gluon distribution 
in the deuteron for two different pomeron gluon distributions: the H1 parametrization
with $\langle x_G \rangle \gg \langle x_v \rangle$ and the prediction from NZ, for which $\langle x_G \rangle \simeq \langle x_v \rangle.$
The dashed lines are for $Q^2=4 GeV^2$, the solid ones for $Q^2=2.25 GeV^2$}
\label{fig3}
\end{figure}
\section{Constraining the glue in the Pomeron by nuclear shadowing}
At present, there are several conflicting claims about the gluon density in the
pomeron: the colour dipole model calculations by Nikolaev and Zakharov, on which our
evaluations of the diffractive NSH are based predicts that gluons and quarks carry
about an equal fraction of the pomeron's momentum, $ \langle x_G \rangle \simeq 
\langle x_v \rangle $. In contrast to that the DGLAP fits to the H1 diffractive 
structure function data prefer the gluon dominated pomeron: $ \langle x_G \rangle \gg 
\langle x_v \rangle $. The two options for glue in the pomeron, $G_\Pom(x,Q^2) =
xg_\Pom (x,Q^2)$ give drastically different predictions for the
shadowing correction
\begin{figure}[h]
\epsfxsize=1.0\hsize
\epsfbox{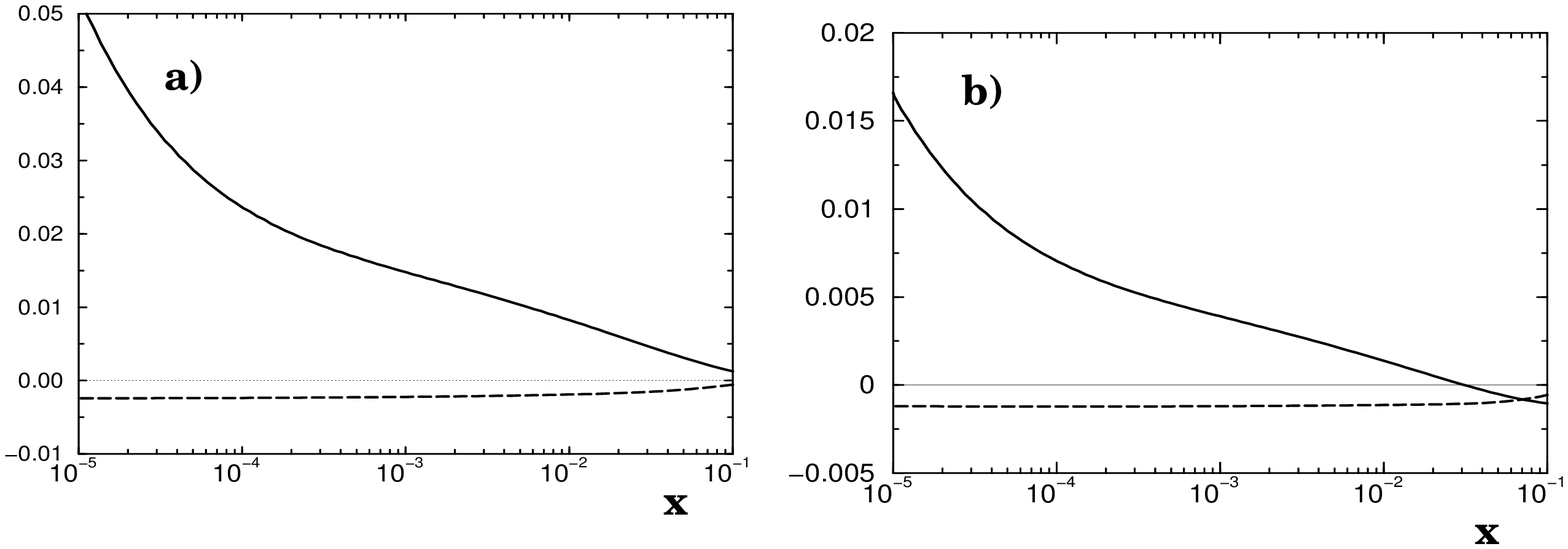}
\caption{(a)Shadowing contributions to the deuteron gluon SF: diffractive NSH(solid line) and
pionic contribution (dashed line).(b)Gluon tensor asymmetry 
$A_G(x,Q^2) = (G^{(\pm)}(x,Q^2)-G^{(0)}(x,Q^2))/G^D(x,Q^2)$: diffractive (solid line) and pionic (dashed line) contributions; $Q^2 = 4GeV^2$.} 
\label{fig3}
\end{figure}
 \be \Delta
G^{(\lambda )}_{sh}(x,Q^2) = \int_x^1 {dx_\Pom \over x_\Pom} \Delta
n_{\Pom/D}^{(\lambda)}(x_\Pom) G_\Pom ({x \over x_\Pom} , Q^2), \ee
to the gluon structure function of the deuteron, see fig 2c). A similar
formula holds for the pionic contribution. There are no direct experimental
data on nuclear shadowing of glue in the deuteron, but precision data
on shadowing in DIS on heavy nuclei allow one to perform the 
DGLAP analysis of nuclear structure functions. Eskola et al. in their recent analysis
find $ R^A_G = \Delta G^{(A)}_{sh}(x,Q_0^2)/AG(x,Q_0^2)
 \simeq 0.15$ at $Q_0^2 \simeq 2.25 GeV^2,
x \simeq 0.001$ for the $^{12}C$--nucleus which is about the same
as shadowing of sea quarks. The straightforward nuclear size 
arguments suggest\cite{Barone} that
$R^A_G(deuteron) \simeq 0.3 \cdot R^A_G(^4 He) \simeq
0.2 \cdot R^A_G(^{12} C)$, i.e. a $\sim 3 {\%}$ effect for the deuteron, which
agrees with the result for the NZ-glue in the pomeron. In contrast, the very strong
shadowing predicted by the H1 glue is in an obvious conflict with the NMC 
experimental data on nuclear shadowing. 
\section{Conclusions}
We have demonstrated that the deuteron tensor structure function at small $x$ is
driven by the inelastic nuclear shadowing. It is rising towards small $x$, which
makes it a unique spin effect -- all other spin effects in total 
cross sections are known to die out at high energies. It was shown that this
fact leads to the conclusion, that the Close-Kumano sum-rule does not exist.
Our findings of a rise of $b_2$ at small $x$ have also been confirmed by 
calculations of other groups \cite{andere}. Further it was demonstrated, that
a hard gluon dominated pomeron structure function leads to strong gluon shadowing 
in the deuteron, which is quite suspect in view of the DGLAP-phenomenology of
nuclear structure functions.


\section*{References}
\begin{thebibliography}{99}
\bibitem{tensor}N.N. Nikolaev and W.Sch\"afer, \Journal{\PLB}{398}{245}{1997}

\bibitem{Hood} P. Hoodbhoy,R.L. Jaffe and A. Manohar, \Journal{\NPB}{312}{571}{1988}

\bibitem{Pais} A. Pais, \Journal{\PRL}{19}{544}{1967} 

\bibitem{Umn} A.Yu. Umnikov \Journal{\PLB}{391}{177}{1997}

\bibitem{Close}F. Close and S. Kumano, \Journal{\PRD}{42}{2377}{1990}.

\bibitem{Gribov}V.N. Gribov, \Journal{\JETP}{29}{483}{1969}.

\bibitem{GNZ}M. Genovese, N.N. Nikolaev and B.G. Zakharov, \Journal{\PLB}{378}{347}{1996}.

\bibitem{NZ1}N.N. Nikolaev and B.G. Zakharov, \Journal{\ZPC}{49}{607}{1991}
  
\bibitem{NZ2}N.N. Nikolaev and B.G. Zakharov, \Journal{\ZPC}{53}{331}{1992}


\bibitem{Barone} V. Barone, M. Genovese, N.N. Nikolaev, E. Predazzi and B.G. Zakharov,
 \Journal{\ZPC}{58}{541}{1993}

\bibitem{Eskola}K.J. Eskola et al., e-print-archives hep-ph/9802350

\bibitem{H1}H1 collab.:C. Adloff et al., \Journal{\ZPC}{76}{613}{1997}

\bibitem{andere} J. Edelmann, G. Piller and W. Weise,  e-print-archives hep-ph/9709455,
K. Bora and R. Jaffe, e-print-archives  hep-ph/9711323

\end{thebibliography}
\end{document}